\documentclass[english]{llncs}
\usepackage[T1]{fontenc}
\usepackage[latin1]{inputenc}
\usepackage{color}
\usepackage{cite}
\usepackage{amstext}
\usepackage{amsmath}
\usepackage{amssymb}
\usepackage[unicode=false,
 bookmarks=false,
 breaklinks=false,colorlinks=false]
 {hyperref}

\usepackage{babel}
\begin{document}

\title{ Statistical distribution of the reversible gates: what percentage
of them are self-inverse? }

\author{Anirban Pathak}

\institute{
Jaypee Institute of Information Technology, A-10, Sector-62,
Noida, India}

\maketitle

\begin{abstract}
It is well known that most of the frequently used reversible
logic gates (e.g., NOT, CNOT, SWAP, Toffoli, Fredkin) are self-inverse
and are represented by square matrices that are unitary and Hermitian.
However, with a simple minded argument, it is established that the
most of the allowed reversible gates are non-self-inverse (unitary
but non-Hermitian) in nature. It is also shown that the \% of non-Hermitian
gates increases with the dimension. For example, 58.33\% of the 2-bit
gates, 98.10\% of the 3-bit gates and 99.99\% of the 4-bit gates are
non-Hermitian. As classical reversible gates are essentially permutation
gates, above statistics is strictly valid for classical reversible
gates, but the argument can be easily extended to include quantum
gates and to establish that the majority of the quantum gates are also
non-self-inverse. Further, the \% of genuinely 2-bit reversible gates
(i.e., 2-bit gates that cannot be decomposed as a product of two single
bit gates) among all possible gates has been computed as 83.3\%, and
the applicability of this analysis in the optimization of circuit
cost is discussed.
\end{abstract}

\section{Introduction}

In 1961, Landauer conjectured that the erasure of information is a
dissipative process which involves a minimal amount of energy loss
\cite{Landauer61}. Thus, thermal energy is produced on erasure of
a classical bit of information. As standard classical gates, such
as NAND, AND, NOR, OR, map 2 bit input states to 1 bit output states
they are irreversible (input state cannot be uniquely reconstructed
from the output state) and involve erasure of 1 bit of information
on every operation of any one of these gates. Consequently, irreversible
gates always involve loss of energy. This is not desirable in today's world,
where a lot of efforts have been made to design low power circuits.
The loss due to the operation of irreversible gates is generally small
compared to the heat loss that happens through the interconnects. However, if we need to
design a truly low power circuit, then we must use reversible gates.
In fact, in a pioneering work, Bennett \cite{Bennett} established
that a (theoretical) zero-power-dissipation circuit must be reversible.
Recently, Landauer's principle has been verified experimentally \cite{experiment1,experiment2}.
These experimental observations have renewed the interest on reversible
circuits, and recently various aspects of reversible circuits have
been investigated. For example, reversible circuits have been designed
for various purposes, including low power encoders \cite{low power}, multipliers \cite{multiplier1,multiplier2}, adder \cite{adder}, GCD finder \cite{GCD}.
A large collection of reversible circuits for various purposes
can be found in benchmark page \cite{benchmarkpage} and Revlib \cite{revlib}. All these reversible circuits
are essentially made of reversible gates and in this paper, we aim
to analyze the statistical distribution of those gates.

A reversible gate is represented by a unitary matrix. This is so because
unitarity ensures that if $U$ exists, then $U^{-1}$ exists. Here
we may note that unitarity demands reversibility, but it does not
demand self reversibility ($U=U^{-1})$. Further, reversibility of
a gate is a well studied domain of research, and there exists a class
of classical reversible gates \cite{Rev-review}. In fact, classical
reversible circuits predate quantum circuits and quantum computing,
which is in much discussion now. However, with the present understanding
of the subject the set of classical reversible gates may be considered
as a subset of the set of all quantum gates with a restriction that
the inputs and outputs in a classical reversible circuit cannot exist
in superposition states. Quantum gates and reversible gates (which
are nothing but  restricted special cases of quantum gates) are the
essential building blocks for implementing various schemes of quantum \cite{dpdincenzo}
and reversible\footnote{When we refer to a gate as reversible gate it implies that the gate
is a classical reversible gate. To specify a quantum gate which is
always reversible we usually refer to the gate as quantum gate.} computation. In what follows, in this paper, we
will discuss classical reversible gates only (unless otherwise stated).
Because of the important role of the reversible gates and circuits,
various aspects of reversible gates and circuits (including synthesis and design mechanisms) have been studied in the recent
past \cite{Rev-review,GCD,multiplier1,multiplier2,adder,benchmarkpage,revlib,withDueck,Dueck1,Dueck2,Dueck3,Wille1,Wille2,Wille3}. However, to the
best  of our knowledge, no effort has yet been made to study the statistical
distribution of the reversible gates. This observation motivated us
to address a few very simple but interesting questions. For example:
What \% of all possible reversible gates is self-inverse? What \%
of all possible two bit gates is genuinely two bit gates or non-decomposable
gates (i.e., cannot be decomposed into single bit gates)? Present
paper aims to address these questions using a very simple, but mathematically
correct logic, which does not involve any approximation. The effort
leads to very interesting conclusions. Specifically, we have observed
that majority of the reversible gates are non-self-inverse. 

The rest of the paper is organized as follows. In Section \ref{sec:Quantum-gates-=000026con-for-self-revers},
we have discussed the condition of self-reversibility and have shown
that self-reversibility implies Hermiticity for unitary operations
(gates). In Section \ref{sec:Statistical-distribution-of}, we have
studied the statistical distribution of the reversible gates. Specifically,
we have reported the \% of non-self-inverse reversible gates and genuinely
two-bit reversible gates. Finally, we conclude the work in Section
\ref{sec:Conclusion}.

\section{Reversible gates and condition for self-reversibility \label{sec:Quantum-gates-=000026con-for-self-revers}}
As quantum gates are more general, we begin by describing the structure of a quantum gate. Reversible gates can be viewed as special cases of quantum gates that does not allow superposition states in input, especially, in the controlled qubit to circumvent the generation of the entangled states (which do not have a classical analogue) in the output.
An $m$-qubit quantum gate is a unitary operator $U$ that maps an
$m$ qubit state into another $m$ qubit state. Since a qubit is a
two level quantum system, in $\{|0\rangle,|1\rangle\}$ basis, we can
write an arbitrary qubit as $|\psi\rangle=\alpha|0\rangle+\beta|1\rangle=\left(\begin{array}{c}
\alpha\\
\beta
\end{array}\right),$ where $|\alpha|^{2}+|\beta|^{2}=1$, and a single qubit gate as a
$2\times2$ unitary matrix. For example, $X=\left(\begin{array}{cc}
0 & 1\\
1 & 0
\end{array}\right)$ is a NOT gate. Similarly, in $2^{m}$ dimensional Hilbert space a
state is a column matrix with $2^{m}$ rows, and an $m$-qubit quantum
gate is a $2^{m}\times2^{m}$ unitary matrix. The general structure of the reversible gates is also the same. Specifically, in the above described general
construction, if we exclude the possibility that the input and output
states can exist in superposition state, then we obtain a reversible
gate. In the previous section, we have briefly mentioned several aspects
of reversible computing that are reported to date. Implementation
of them requires reversible gates. The point we wish to establish here
is that most of the reversible gates (unitary operators) are non-self-inverse.
To elaborate the idea we will start with some features of reversible
gates. 

Here, we would like to note an interesting feature of reversible gates:
A unitary operator $A$ must satisfy $A^{-1}=A^{\dagger}$ and a Hermitian
operator $A$ must satisfy $A=A^{\dagger}.$ Consequently, all unitary
operators are not Hermitian and all Hermitian operators are not unitary.
If a unitary operator $A$ is found to be Hermitian, then $A^{-1}=A^{\dagger}=A$,
i.e., $A=A^{-1}$, so the operator is self-inverse. We can easily
show the converse (i.e., self inverse unitary operators are Hermitian).
Thus, all the reversible gates that are represented by Hermitian unitary
matrices are self-inverse and other reversible gates (i.e., those
gates that are represented by non-Hermitian unitary matrices) are
non-self-inverse. For example, we can easily observe that ${\rm {\rm CNOT}}$
gate $|00\rangle\langle00|+|01\rangle\langle01|+|11\rangle\langle10|+|10\rangle\langle11|$
is self-inverse, naturally it is Hermitian, too. Now, we can drop
the symmetry required for the gate to be self-inverse and modify it
to another quantum gate $B=|01\rangle\langle00|+|11\rangle\langle01|+|00\rangle\langle10|+|10\rangle\langle11|$.
Clearly this gate is not self-inverse as $BB|00\rangle=B|01\rangle=|11\rangle\neq|00\rangle$.
Thus, this gate is non-Hermitian, but interestingly it is unitary.
So it is an example of non-self-inverse reversible gate. Now, we may
note that all the popularly used reversible gates, like NOT, CNOT,
Toffoli, SWAP, Fredkin, are self-inverse, but non-self-inverse reversible
gates, like B, do exist. This observation leads to a question: What
\% of the reversible gates are self-inverse? In the next section,
we aim to address this question.

\section{Statistical distributions of reversible gates\label{sec:Statistical-distribution-of}}

Assume that we are working in an $M$ dimensional Hilbert space and
the input states are $\{|a_{1}\rangle,|a_{2}\rangle,|a_{3}\rangle,\cdots,|a_{M}\rangle\}$.
Thus, $\left\{ |a_{i}\rangle\right\} $ forms our input basis set.
Similarly, assume that $\{|b_{j}\rangle\}$ represents a new basis
set in the same dimension, and it is our output basis set. Now we
may introduce the operators $U_{J}=\sum_{j}|b_{j}\rangle\langle a_{j}|$,
which are unitary, as is easily verified to satisfy $U_{J}U_{J}^{\dagger}=U_{J}^{\dagger}U_{J}=\left(\sum_{p}|a_{p}\rangle\langle b_{p}|\right)\left(\sum_{q}|b_{q}\rangle\langle a_{q}|\right)=\left(\sum_{j}|a_{j}\rangle\langle a_{j}|\right)=\mathcal{I}_{M}$.
Let $\Pi_{J}$ ($J=1,2,\cdots,M!$) be an arbitrary permutation on
$M$ letters, each of this permutation will provide us a unitary operator.
Each of these $M!$ unitary operators $U_{J}$ is a reversible gate
that always maps input states into mutually orthogonal output states.
We may refer to these reversible gates as permutation gates, but these
unitary gates are not essentially self-inverse (Hermitian). This fact
is elaborated through a specific example in the previous section,
where the operator $B$ was unitary but non-self-inverse. Here, we
would like to note that in general (say when you include quantum gates,
too) $\left\{ |b_{j}\rangle\right\} $ is only required to be a basis
set in $M$ dimension, and it is not required to be a permutation
of $\left\{ |a_{i}\rangle\right\} $. However, for reversible gates
$\left\{ |b_{j}\rangle\right\} $ is only permutation of $\left\{ |a_{i}\rangle\right\} $
as superposition states are not allowed.  For each unique choice of
input basis $\left\{ |a_{i}\rangle\right\} $, we can construct $M!$
reversible gates in $M$ dimension and each of these gates can map
input states to mutually orthogonal output states. In essence, these
are permutation gates.

Permutation gates introduced above can be expressed in a compact notation
as follows: 
\begin{equation}
U=\left(\begin{array}{cccc}
|a_{1}\rangle & |a_{2}\rangle & \cdots & |a_{M}\rangle\\
|b_{k}\rangle & |b_{l}\rangle & \cdots & |b_{n}\rangle
\end{array}\right)=\left(\begin{array}{cccc}
1 & 2 & \cdots & M\\
k & l & \cdots & n
\end{array}\right)=(k,l,\cdots,n).\label{eq:notation}
\end{equation}
This notation explicitly identifies the positions occupied by elements
before and after application of a permutation gate. The notation uses
a matrix, where the states in first row are in $\left\{ |a_{i}\rangle\right\} $,
and those in the second row are the new arrangement where the states
are in $\left\{ |b_{i}\rangle\right\} $. This notation is familiar.
Now, for our convenience, we introduce another notation to specify
this permutation gate. In our notation, numbers in a row indicate
new positions of the state vectors. For example, $(1,2,3,\cdots,M)$
will indicate identity gate, $(2,1,3,4,5,\cdots,M)$ will denote a
specific permutation gate that maps $|a_{2}\rangle$ to $|a_{1}\rangle,$
$|a_{1}\rangle$ to $|a_{2}\rangle$ and $|a_{i\neq1,2}\rangle$ to
$|a_{i}\rangle$. Clearly, $(2,1,3,4,5,\cdots,M)$ is a self-inverse
gate and thus Hermitian gate. Further, in our compact notation if
both input basis and output basis are computational basis (this is the case, we should consider for all the classical reversible gates) then $I_{2}=\left(\begin{array}{cc}
|0\rangle & |1\rangle\\
|0\rangle & |1\rangle
\end{array}\right)=(1,2)$.

\subsection{What \% of all possible reversible gates is non-self-inverse?}

In this section, we aim to answer, how many of $M!$ permutation gates
obtained for a specific choice of input basis $\left\{ |a_{i}\rangle\right\} $
and output basis $\{|b_{j}\rangle\}$ are self-inverse? As in this
context (when we know that the operator/gate is unitary) Hermiticity
is equivalent to self-inverse. Now Identity gate is always self-inverse.
In 2 dimension, we can choose two state vectors in $^{2}C_{2}=1$
way and interchange them to form a self-inverse gate. In 2 dimension,
there is only one permutation gate except $(1,2)$, i.e., $(2,1)$
and that is self-inverse (it is NOT gate if the input basis $\left\{ |a_{i}\rangle\right\} =\{|b_{j}\rangle\}$.
Now in 4 dimension, we may choose 2 state vectors in $^{4}C_{2}=6$
ways and form the following Hermitian gates $(2,1,3,4),\,(3,2,1,4),\,(4,2,3,1),\,(1,3,2,4),$ $(1,4,3,2),$
and $(1,2,4,3)$. In addition, Identity is a Hermitian gate. Further,
we can choose 4 state vectors in pairs of 2 in $\frac{^{4}C_{2}}{2}=3$
ways and form the following 3 self-inverse gates $(2,1,4,3),\,(3,4,1,2),\,(4,3,2,1)$.
Thus, in total there are $6+3+1=10$ self-inverse (Hermitian) permutation
gates in $C^{4},$ which includes ${\rm CNOT}=(1,2,4,3)$ gate. Remaining
$4!-10=14$ permutation gates are unitary but non-Hermitian. They
are $(2,3,1,4),$ $(2,3,4,1),$ $(1,3,4,2),$ $(1,4,2,3),$ $(2,4,1,3),$
$(2,4,3,1),$ $(3,1,2,4),$ $(3,1,4,2),$ $(3,2,4,1),$ $(3,4,2,1),$ $(4,1,2,3),$ 
$(4,1,$ $3,2),$ $(4,2,1,3),$ and $(4,3,1,2)$. Interestingly, the number
of possible non-Hermitian permutation gates in $C^{4}$ are more than
the possible Hermitian gates in $C^{4}$. To have a quantitative perception
we may note that in $C^{2^{2}}$, $\frac{14}{24}\times100\%=58.33\%$
of the possible permutation gates are non-Hermitian. Now we need to
generalize this idea to $C^{2^{n}}.$ In order to do so we have to
find out the number of self-inverse permutations on $M$ letters,
also known as involutions which is denoted as $a[M]$. Involutions
or the number of alternative permutations are a well studied topic in mathematics.
For $M=1,2,3,\cdots$ number of alternating permutations are $a[M]=1,2,4,10,26,76,232,764,$
2620, 9496, 35696, 140152, 568504, 2390480, 10349536, 46206736, 211799312,
...{[}See Ref. \cite{involution} and references therein{]}. Thus,
in the context of the present work, the number of $n$-bit
self-inverse gates for a specific choice of input bases is $a[2^{n}]$
and consequently \% of non-self-inverse $n$-bit
reversible gates for input basis set is\footnote{It is easy to obtain $r(n)$ using a small Mathematica program: $a[\text{n\ensuremath{\_}}]=\text{If}[n<0,0,n!\text{SeriesCoefficient}[\text{Exp}[x+x{}^{\wedge}2/2],\{x,0,n\}]];r[\text{n\ensuremath{\_}}]=\frac{2^{n}!-a\left[2^{n}\right]}{2^{n}!}100.0.$
Here, the expression for $a[n]$ is used from Ref. \cite{involution}.} $r(n)=\frac{\left(2^{n}\right)!-a[2^{n}]}{\left(2^{n}\right)!}\times100\%.$
 Now we can easily check that $r(3)=98.1052,\,r(4)=99.9998$. Therefore,
98.10\% of the 3-bit reversible gates and 99.999\%
of the 4-bit reversible gates are non-Hermitian. It is interesting
to note that most of the possible permutation gates are non-Hermitian
and their \% increases with the increase in dimension. This is more
interesting because most of the popularly used reversible gates (e.g.,
NOT, CNOT, SWAP, Toffoli, Fredkin gates) are self-inverse.

\subsection{What \% of all possible 2-bit reversible gates are
non-decomposable gates?}

A 2-bit reversible gate that cannot be constructed
as a tensor product of two single bit gates is known
as non-decomposable gate. Since such a reversible gate cannot be decomposed
into single bit gates, it is considered as genuinely 2-bit
reversible gate. In other words, the non-decomposable gates are the
actual 2-bit gates. To be precise, a unitary gate
acting on a bipartite system $A\otimes B$ is called decomposable 
if $U=U_{A}\otimes U_{B},$ where $U_{A}$ and $U_{B}$ are single
bit gates. If $U$ cannot be expressed as $U=U_{A}\otimes U_{B}$
then the gate is considered to be non-decomposable. Non-decomposable
2-bit gates are very important for reversible circuit synthesis and optimization of quantum cost of the reversible circuits \cite{withDueck}. This fact motivates us to ask a question:
How frequent are the non-decomposable 2-bit gates?

A 2-bit identity gate is a product of two single bit identity gates.
Apart from this trivial example, there are three more decomposable
for the allowed choice of input and output basis sets. They are (i)
$(1,2)\otimes(3,4)=(3,4)\otimes(1,2),$ (ii) $(1,2)\otimes(4,3)=(4,3)\otimes(1,2)$
and (iii) $(2,1)\otimes(3,4)=(2,1)\otimes(4,3).$ Thus, 4 out of 24
possible reversible 2-bit gates are decomposable. In
other words, $\frac{20}{24}\times100\%=83.3\%$ of the 2-bit
gates are non-decomposable gates or genuinely 2-bit
gates. Thus, among all possible 2-bit permutation gates, non-decomposable
gates are more common than the decomposable gates.

\section{Conclusion\label{sec:Conclusion}}

Various aspects of reversible gates have been studied for long.
However, no effort has yet been made to analyze the statistical distribution
of the reversible gates from any perspective. The present work provides a
fair idea of statistical distribution of reversible gates with respect
to (i) self-inverse and non-self-inverse gates and (ii) decomposable
and non-decomposable 2-bit reversible gates. The outcome
of the analysis is interesting from purely foundational as well as
application perspective. Specifically, present approach provides some
useful idea to design new templates which are expected to be useful
in optimizing reversible circuits. To be precise, a template is a
reversible circuit equivalent to Identity. Templates are known to
be useful for optimization of reversible circuits \cite{template}.
As in the present approach the gates obtained using a fixed input
and output bases forms a symmetric group $S_{N},$ we can use the
group multiplication table to form new templates. For example, if
there are $N$ gates we have inverse of all of them in the set so
rearrangement theorem tells us that we have $N$ templates of two
gates. Now if we wish to convert a particular template $U_{1}U_{2}=I$
into a three gate template, we can keep $U_{1}$ fixed and decompose
$U_{2}$ into two gates and the group rearrangement theorem tells
us that the same can be done in $N$ ways, so we have $N$ templates
of the form $U_{1}U_{i}U_{j}$. In a similar manner, we may generate
many simple templates which may be found useful in the optimization of
quantum circuits.

We often construct reversible circuits using NCT gate library \cite{benchmarkpage},
all the gates of this library and the other frequently used reversible
gates like Fredkin gate are self-inverse. However, this situation
is really special as it is rare. Specifically, we have seen that 58.33\%
of the 2-bit reversible gates, 98.10\% of the 3-bit
reversible gates and 99.99\% of the 4-bit reversible gates are non-self-inverse.

\textbf{Acknowledgment:} Author thanks Department of Science and Technology (DST), India for the support
provided through the project number EMR/2015/000393. He also thanks Dr. A. Banerjee and Mr. K. Thapliyal for their interest
in the work and for some useful technical discussions.

\end{document}